\documentstyle[prl,aps,multicol]{revtex}

\newcommand{\be}{\begin{eqnarray}}
\newcommand{\ee}{\end{eqnarray}}
\newcommand{\non}{\nonumber\\}
\newcommand{\inline}[1]{\noalign{\hbox{#1}}}
\newcommand{\ave}[1]{\left\langle #1 \right\rangle}


\begin{document}

\title{Charged Particle Ratio Fluctuation as a Signal for QGP}

\author{ S. Jeon\thanks{ e-mail: sjeon@lbl.gov } and 
V. Koch\thanks{ e-mail: vkoch@lbl.gov } }

\address{Nuclear Science Division\\
Lawrence Berkeley National Laboratory\\
Berkeley, CA 94720, USA}

\maketitle

\begin{abstract}
In this letter we argue that the event-by-event fluctuations of the ratio
of the positively charged and the negatively charged
pions provides a signal of quark-gluon plasma.
The fact that quarks carry fractional charges is ultimately responsible
for this distinct signal.

\end{abstract}

\begin{multicols}{2}

 It is of great importance that we have a clear signal of the
 long-sought quark-gluon plasma (QGP) not only for the experiments at RHIC
 but also for theoretical reasons.  
 At stake is our fundamental understanding of strong interactions 
 as well as understanding of the state of matter 
 in the very early universe\cite{Bonometto:1993pj}.
 Proposed signals of this new state of matter abound in literature
 \cite{Blaizot:1999bv}
 one of
 the most studied being the $J/\psi$ suppression 
 \cite{Satz:1998bd,Gale:1999ve}.
 
 In this paper, we propose the event-by-event $h^+/h^-$ fluctuations as
 a distinct signal of QGP formation.
 We would also like to stress that this observable is something that can
 be and already has been calculated on a lattice.
 
 The idea is very simple and is reminiscent of the original detection of
 color in $e^+ e^-$ experiment where one measures
 \be
 R_{e^+e^-} \equiv
 {e^+e^- \to \ \hbox{Hadrons}
 \over
 e^+e^- \to \mu^+ \mu^-} 
 =
 N_c \sum_{q} Q_{q}^2
 \ee
 Here $Q_q$ is the charge of each flavor and $N_c$ is the number of
 colors.
 Note that if the fundamental degrees of freedom were hadrons, 
 $R_{e^+e^-}$ would be very different from this simple counting.
 We would like to establish that the event-by-event $h^+/h^-$ fluctuations
 can similarly determine whether the underlying
 degrees of freedom are quarks and gluons or hadrons.

 The point is that in the QGP phase, the unit of charge is $1/3$
 while in the hadronic phase, the unit of charge is 1.
 The net charge, of course does not depend on such subtleties.
 However, the fluctuation in the net charge depends on 
 the {\em squares} of the charges and hence strongly depend 
 on which phase it originates from.  
 Measuring the charge fluctuation itself, however, is plagued by
 systematic uncertainties such as volume fluctuations due to
 the impact parameter variation. 
 In a previous letter\cite{Jeon:1999gr}, we showed that the multiplicity
 ratio fluctuation is only sensitive to the {\em density} fluctuations
 and not to the volume fluctuations. 
 The task for us is then to find a suitable ratio whose fluctuation 
 is easy to measure and simply related to the net charge fluctuation. 
 
 The obvious candidate is the ratio $F = Q/N_{\rm ch}$ where
 \be
 Q = N_+ - N_-
 \ee
 is the net charge and 
 \be
 N_{\rm ch} = N_+ + N_-
 \ee
 is the charge multiplicity. Here $N_\pm$ denote the positive and
 negative multiplicities.
 Instead of using $F$, however, in this paper we propose to use 
 the charge ratio $R = N_+/N_-$. 
 The advantages of using $R$ over $F$ are that although trivially
 related, $R$ is more fundamental to experiments 
 and the signal is about 4 times amplified in $R$ as we show below.
 
 To relate $R$ with $F$,
 we first rewrite the charge ratio as
 \be
 R 
 = 
 {N_+ \over N_-} 
 =
 {1 + F \over 1 - F}
 \ee
 When $\ave{N_{\rm ch}} \gg \ave{Q}$
 we can safely say $\left| F \right| \ll 1$.
 Expanding in terms of $F$ yields
 \be
 & R \approx 1 + 2F + 2F^2 &
 \ee
 Defining $\delta x = x - \ave{x}$ for any fluctuating quantity $x$,
 it is easy to show 
 \be
 \ave{\delta R^2}
 = 
 \ave{R^2} - \ave{R}^2
 \approx  
 4\ave{\delta F^2}
 \ee
 where $\langle \cdots \rangle$ denotes the average over all events.
 
 Let us now consider $\ave{\delta F^2}$ more closely.  
 In a previous letter \cite{Jeon:1999gr}
 (see also \cite{Baym:1999up}
  and the upcoming paper \cite{BJK}), 
 we showed that a ratio fluctuation can be expressed as
 \be
 \ave{\delta F^2}
 =
 {\ave{Q}^2\over \ave{N_{\rm ch}}^2}
 \ave{
 \left(
 {\delta Q\over {\ave Q}}
 -
 {\delta N_{\rm ch}\over {\ave N_{\rm ch}}}
 \right)^2
 }
 \ee
 We then showed that when the average ratio is very
 much different than 1, the fluctuation is driven mainly by the
 fluctuation in the smaller quantity (for instance $K/\pi$ fluctuation
 is driven by $K$ fluctuation).
 At RHIC we expect $\ave{Q}/\ave{N_{\rm ch}} \sim 5 \%$.
 Hence the fluctuation in $F$ 
 is totally dominated by the fluctuation in $Q$ so that
 \be
 \ave{\delta F^2} \approx {\ave{\delta Q^2}\over \ave{N_{\rm ch}}^2}
 \ee
 If we can detect all charged particles from a heavy ion collision, 
 the net charge $Q$ is a fixed quantity and hence will not fluctuate.
 This implies that $\ave{\delta F^2}$ is very small with a $4\pi$
 coverage.
 However, no detector can catch all charged particles.
 Our study \cite{Jeon:1999gr,BJK} shows that for a realistic
 detector acceptance, using the grand canonical ensemble is acceptable and
 that is what we assume here.
 The corrections to this approximation have been  
 worked out and will be reported in the upcoming paper\cite{BJK}.  
 (Also see {\bf Finite Acceptance Correction} at the end of this
 letter.)
 Our main observable is then, 
 \be 
 \ave{N_{\rm ch}}\ave{\delta R^2} = 4\ave{N_{\rm ch}}\ave{\delta F^2} 
 =4 {\ave{\delta Q^2} \over \ave{N_{\rm ch}}}
 \label{eq:Dpm}
 \ee
 to the leading order in the fluctuations and $1/\ave{N_{\rm ch}}$. 
 
 So far, we have only considered statistics of the ratio fluctuations.  
 Physics lies in how the charge fluctuation is expressed in terms of the
 fluctuations in the fundamental degrees of freedom.
 For simplicity, let us consider a pion gas and a QGP consisting of $u$
 and $d$ quarks and gluons.  Our main conclusion does not depend
 on this simplifying assumption.  We will briefly consider 
 the size of the corrections towards the end of the paper. 
 
 In a pion gas, the fundamental degrees of freedom are of course pions.
 Hence, $N_{\rm ch} = N_{\pi^+} + N_{\pi^-}$ and
 \be
 \delta Q = \delta N_{\pi^+} - \delta N_{\pi^-}
 \ee
 Using thermal distributions and disregarding correlations, we get 
 \be
 \ave{\delta Q^2} = \ave{\delta N_+^2} + \ave{\delta N_-^2}
 =
 w_\pi\ave{N_{\rm ch}}
 \label{eq:numerator}
 \ee
 where
 \be
 w_\pi \equiv \ave{\delta N_\pi^2}/\ave{N_\pi}
 \label{eq:wpi}
 \ee
 is slightly bigger than 1 \cite{Jeon:1999gr,Baym:1999up}.
 Hence for a pion gas, 
 \be
 D_{\rm had} 
 \equiv \left.\ave{N_{\rm ch}}\ave{\delta R^2}\right|_{\rm hadron}
 \approx 4
 \;.
 \label{eq:pion_gas}
 \ee
 
 For a quark-gluon plasma, 
 \be
 \delta Q = Q_u\, \delta\left( N_u - N_{\bar u}\right)
          + Q_d\, \delta\left( N_d - N_{\bar d}\right)
 \ee
 where $Q_{q}$ is the charges of the quarks 
 and $N_{q}$ is the number of quarks.
 The fluctuations $\ave{\delta N_{u,d}^2}$ are
 measured on lattice\cite{Gottlieb:1997ae} and we will shortly get back
 to the results.  For now, let us consider a thermalized gas of
 non-interacting quarks and gluons to get a physical baseline.
 Thermal distributions and no correlations yield
 \be
 \ave{\delta Q^2} 
 = Q_u^2 w_u \ave{N_{u+\bar u}} 
 + Q_d^2 w_d \ave{N_{d+\bar d}}
 \label{eq:dQ2}
 \ee
 where $N_{q+\bar q}$ from here on denotes the number of 
 quarks {\em and} anti-quarks. 
 The constant
 \be
 w_q \equiv \ave{\delta N_q^2}/\ave{N_q}
 \label{eq:wq}
 \ee
 is slightly smaller than 1 due to the fermionic nature of the quarks.

 Relating the final charged particle multiplicity $N_{\rm ch}$ to the
 number of primordial quarks and gluons is not as simple.
 To make an estimate, we assume that the entropy is conserved 
 \cite{Bjorken:1983qr}
 and that all the particles involved are
 massless, in thermal equilibrium and non-interacting.
 For such particles, the following relation between the
 entropy density and the particle number density holds:
 \be
 \sigma_B = 3.6 n_B
 \\
 \inline{and}
 \sigma_F = 4.2 n_F
 \ee
 where the subscript $B, F$ signifies the particle types.
 The total entropy of a quark-gluon gas in a given volume $V_{qg}$ is
 \be
 S = V_{qg} \sigma_{qg} 
 = 3.6 \ave{N_g} 
 + 4.2\left( \ave{N_{u+\bar u}} + \ave{N_{d+\bar d}} \right)
 \ee
 where $N_{g}$ is the number of gluons inside the volume and
 $N_{q+\bar q}$ is
 the number of quarks and anti-quarks inside the same volume.
 As the volume expands and cools, eventually the quarks and gluons are
 converted to pions.  Since entropy is conserved, the number of
 pions coming from these quarks and gluons must be given by
 \be
 \ave{N_\pi} = {S\over 3.6} = 
 \ave{N_g} + {4.2\over 3.6}
 \left(\ave{N_{u+\bar u}} + \ave{N_{d+\bar d}}\right)
 \label{eq:entropy}
 \ee
 The charged multiplicity is $2/3$ of $N_\pi$ due to isospin symmetry:
 \be
 \ave{N_{\rm ch}}
 = {2\over 3}
 \left(\ave{N_g} + 1.2 \ave{N_{u+\bar u}} + 1.2 \ave{N_{d+\bar d}} \right)
 \label{eq:Nch}
 \ee
 Then for massless non-interacting quarks and gluons, 
 \be
 D_{\rm QGP} 
 \equiv \left. \ave{N_{\rm ch}}\ave{\delta R^2} \right|_{\rm QGP}
 \approx 0.75 
 \label{eq:qgp}
 \ee
 from Eq.(\ref{eq:Dpm}) and using Eqs.(\ref{eq:dQ2}) and (\ref{eq:Nch}). 
 The value of $D_{\rm QGP}$ is more than a factor of 5 smaller than 
 the value of $D_{\rm had}$ in Eq.(\ref{eq:pion_gas})!
 This is an unmistakable signal of QGP formation from such a simple 
 measurement.   

 We now would like to stress that 
 $\ave{\delta Q^2}/\ave{N_{\rm ch}}$ is already calculated on lattice
 and hence one does not have to rely on the above thermal model
 calculation.
 In Ref.\cite{Gottlieb:1997ae},
 Gottlieb et~al report their calculation of
 the quark number susceptibility and the entropy density 
 with 2 flavors of dynamic quarks.
 These two quantities are directly related to the net charge fluctuation
 and the charged multiplicity in the following way.
 
 From the definition of the charge susceptibility $\chi_{q}$,
 it is clear that
 \be
 \ave{\delta Q^2} = V_{qg}\, T \chi_q
 \ee
 where $V_{qg}$ is the volume of the quark-gluon plasma at the
 hadronization and $T$ is the temperature.
 Gottlieb et~al calculated  
 the quark number density susceptibilities
 \be
 T\chi_S = 
 V_{qg}\ave{\left(\delta n_u + \delta n_d\right)^2}
 \\
 \inline{and}
 T\chi_{NS} = 
 V_{qg}\ave{\left(\delta n_u - \delta n_d\right)^2}
 \ee
 and found that at high temperature both are very close to
 the non-interacting thermal gas limit
 \be
 \chi_S \approx \chi_{NS} \approx 2 T^2
 \ee
 From this result, one can first of all infer that $u$ and $d$ quark
 densities are uncorrelated
 \be
 &
 \ave{\delta N_u \delta N_d} \approx 0
 &
 \\
 \inline{and}
 & 
 \ave{\delta N_u^2} \approx \ave{\delta N_d^2} 
 \;.
 &
 \ee
 These results imply that the charge fluctuation
 at high temperature follows that of the thermal fermion gas 
 \be
 & 
 \ave{\delta Q^2} 
 =  
 {4\over 9}\ave{\delta N_u^2} 
 +
 {1\over 9}\ave{\delta N_d^2} 
 = {5\over 9} V_{qg} T^3 
 &
 \\
 \inline{or}
 &
 T\chi_q = {5\over 9}T^3
 &
 \;.
 \label{eq:Tchiq}
 \ee

 For the charged multiplicity, 
 we assume that the relations Eqs.~(\ref{eq:entropy} -- \ref{eq:Nch})
 still hold.
 Equating the entropy of the final pions with that of the primordial
 quarks and gluons, one obtains
 \be
 \ave{N_{\rm ch}} = {1\over 5.4} V_{qg} \ave{\sigma_{qg}}
 \;.
 \label{eq:Nchlattice}
 \ee
 Ref.\cite{Gottlieb:1997ae} 
 reports that the gluon entropy density $\sigma_{g}$
 is almost the same as the non-interacting thermal bosons, but the quark
 entropy density $\sigma_{u+d}$ is about one half of that of 
 the non-interacting thermal fermions. 
 Hence, the total entropy from the lattice calculation is
 \be
 \ave{\sigma_{qg}}
 & = &
 \ave{\sigma_g} + \ave{\sigma_{u+d}}
 \non
 & = &
 16\times 3.6 \times \ave{f_g} 
 +
 24\times 4.2 \times \alpha \ave{f_q}
 \non 
 & \approx &
 12 \, T^3
 \label{eq:sigmaqg}
 \ee
 where $\alpha \approx 1/2$
 and $\ave{f_{q,g}}$ is the average density per degree of freedom.
 
 Using Eqs.(\ref{eq:Tchiq} -- \ref{eq:sigmaqg}), the lattice calculation
 gives 
 \be
 D_{\rm lat} 
 \equiv \left. \ave{N_{\rm ch}}\ave{\delta R^2} \right|_{\rm lattice}
 \approx 1
 \label{eq:lattice_result}
 \ee
 which is still 4 times smaller than the pion gas result.
 This is the main result of this paper.  The difference between the pion
 gas result and the QGP result is distinct enough one should easily see
 it in the first few days of data collecting at RHIC. 
 We are of course aware that lattice result is not yet
 exact.  We hope that this discussion will actually stimulate more
 sophisticated lattice calculations of quark number susceptibilities.
 We also note that this is an opportunity to test what is calculated 
 on lattice in a direct observation.

 The picture obtained above holds if the following two conditions are
 met: (i) The detected phase space is a small sub-system of the whole.
 (ii) The original quarks and gluons stay in the system 
 during or after the hadronization.
 Both conditions can be met if 
 the rapidity intervals are such that
 \be
 y_{\rm total} \gg y_{\rm accept} \gg 1
 \label{eq:validity}
 \;.
 \ee
 Here, $y_{\rm total}$ is the rapidity range allowed by the energetics
 of the collisions and $y_{\rm accept}$ is the acceptance interval
 of a given detector.
 The first of these conditions is needed to ensure that the rest of the
 system acts as a reservoir and the second condition ensures that the
 charge diffusion in the rapidity space during and after hadronization 
 is negligible.
 In real life, of course, Eq.~(\ref{eq:validity}) is satisfied in
 varying degrees.  For instance, the STAR at RHIC has
 $y_{\rm total} \approx 10$, $y_{\rm accept} \approx 3$ and hence
 corrections should be taken into account.
 We would like now to discuss Caveats and corrections due to these and
 other effects.
 
 \noindent
 {\bf Hadronization :} 
 First, charge conservation fixes the net charge once it is set in the
 QGP phase.  Subsequent hadronization cannot change the net charge. 
 Second, the entropy can only increase during the hadronization. 
 Hence, our estimate of $\ave{\delta Q^2}/\ave{N_{\rm ch}}$ from QGP 
 should not only survive the hadronization but also could even become 
 smaller, thus strengthening the signal. 
 
 \noindent
 {\bf Phase Space Cuts :}
 A possible issue of having a rapidity cut 
 is that this implies a box in the momentum space
 while the argument presented so far dealt with a box in the coordinate
 space.  Within the Bjorken scenario, this is of course not a problem. 
 However this is not a problem even without the Bjorken picture.  
 If anything, it actually helps.  
 Once it is established that the ratio $h^+/h^-$
 is independent of the overall volume, the only crucial
 ingredient in our argument
 is that the number fluctuations are all Poissonian, namely,
 \[
 \ave{\delta N^2}/\ave{N} \approx 1
 \]
 The quantum statistics makes this ratio deviate from 1.
 For bosons (pions), restricting to small momenta or rapidity
 makes this ratio bigger.
 For fermions (quarks), it makes this ratio smaller. 
 From Eqs.(\ref{eq:numerator}), (\ref{eq:wpi}), (\ref{eq:dQ2}) and
 (\ref{eq:wq}), it is clear that having a momentum cut-off
 can then only enhance the contrast. 
 
 \noindent
 {\bf Resonance Contributions :}
 As explained in a previous paper\cite{Jeon:1999gr}, 
 neutral resonances introduce positive correlations between 
 $N_+$ and $N_-$ and hence lower the value of 
 $D_{\rm had}$ from 4.  In a thermal scenario studied in the
 same paper, we found that the resonances reduce the fluctuation by
 about 30 \%.  Hence for a realistic hadron gas, 
 \be
 D_{\rm had} = \ave{N_{\rm ch}}\ave{\delta R^2} \approx 3
 \;.
 \label{eq:hadron_gas}
 \ee
 This is still a factor of 3 bigger than the lattice result.

 \noindent
 {\bf Mixtures :}
 If the system is a mixture of a QGP and a hadron gas, the signal should
 depend on the fractions.  To a first approximation, it should be a
 linear combination of 
 Eqs.~(\ref{eq:hadron_gas}) and (\ref{eq:lattice_result})
 \[
 \ave{N_{\rm ch}}\ave{\delta R^2} 
 =
 D_{\rm had}(1-f) + D_{\rm lat} f 
 \]
 where $f$ is the QGP fraction.  Even if $f \sim 0.5$, the signal can 
 be still visible.
 
 \noindent
 {\bf Rapidity Correlations :} 
 If gluon interactions dominate the creation of hadrons in high energy
 hadron-hadron collisions, the unlike charged particles are
 strongly correlated in rapidity.  Such strong correlation 
 can further lower the value of $D_{\rm had}$.
 If the reduction is big enough to mimic QGP signal,
 we should certainly see the false signal in even in
 the nucleon-nucleon collisions. 
 Using the two particle correlation data in $pp$ collisions compiled 
 by Whitmore\cite{Whitmore:1976ip}, we performed Monte Carlo simulations
 for such a scenario.  The correction to (\ref{eq:hadron_gas})
 is found to be small ($\sim 10~\%$).  This is due to the 
 {\em experimental fact} that even
 though the correlation is pronounced in the `connected' part of the
 correlation $R(y_+, y_-)$, it is not so pronounced in the full two
 particle correlation function
 \be
 \rho_2(y_+, y_-)
 =
 \rho_1(y_+)\rho_1(y_-)\,[1 + R(y_+, y_-)]
 \;.
 \ee
 For details, see~\cite{BJK}.

 \noindent
 {\bf Finite Acceptance Correction :} 
 The finite size of the acceptance window 
 introduces a factor of $(1-p)$ corrections where 
 $p$ is the fraction of the total multiplicity inside the acceptance window.
 This is easy to understand.  If the detector sees a 100 \% 
 of all charged particles ($p=1$), 
 the fluctuations should shrink down to zero due to the
 global charge conservation.
 Fortunately, this is a common factor that applies to both
 Eqs.~(\ref{eq:hadron_gas}) and (\ref{eq:lattice_result})
 so that the ratio stays the same.
 More details will be reported in~\cite{BJK}.
 
 \noindent
 {\bf Effects of Rescattering :}
 The partons as well as hadrons are subject to rescattering during the
 course of a heavy ion collision.  This in principle may affect the  
 above charge fluctuations by diffusing the charge in rapidity space.
 However in the limit $y_{\rm total} \gg y_{\rm accept} \gg 1$
 these effects should be very small since they scale as the surface to
 volume ratio in rapidity space.
 To estimate the effect in the hadronic phase, 
 we performed a simple Monte Carlo calculation
 where a Gaussian noise with $\sigma = 0.5$ are added to each particle's
 rapidity originating from a highly correlated source.  Doing so increases
 the value of $\ave{N_{\rm ch}}\ave{\delta R^2}$ up to 40 \% assuming the
 STAR acceptance at RHIC.    
 In a slightly different context, Gavin \cite{Gavin:1999bk} also found
 that when a strong flow is established, the rescattering effect on the
 hadron rapidities is relatively small.
 
 \noindent
 {\bf Strangeness :}
 Adding Kaons to a pion gas will not change the value of 
 $D_{\rm had}$ because their contribution adds
 exactly the same amount to both numerator (c.f. Eq.(\ref{eq:numerator}))
 and the denominator.
 For a quark gluon plasma,
 the lattice calculation 
 \cite{Gottlieb:1997ae} 
 suggests that at high temperature,
 the strangeness entropy is about 40 \% of the $u+d$ entropy.
 Taking this at a face value changes the result (\ref{eq:lattice_result})
 by less than 10 \%. 

 In conclusion, we showed in this paper that 
 detection of QGP formation is quite possible through 
 the simple measurement of $h^+/h^-$ fluctuation. 
 This measurement should be very feasible for STAR.
 We also emphasize that this is a direct confirmation of the 
 lattice QCD results.
 
 What we considered here is the simplest ratio out of many possible ones
 that can behave quite differently in the presence of a QGP. 
 For instance, the strangeness anti-strangeness ratio
 fluctuations can provide us with a valuable handle on the strangeness
 distributions with or without the formation of a QGP. 
 These and related issues are under active investigation and will be
 reported elsewhere \cite{BJK}.

 S.J.~would like to thank M.~Bleicher for many discussions.
This work was supported by the Director, 
Office of Energy Research, Office of High Energy and Nuclear Physics, 
Division of Nuclear Physics, and by the Office of Basic Energy
Sciences, Division of Nuclear Sciences, of the U.S. Department of Energy 
under Contract No. DE-AC03-76SF00098.

{\it Note added:}
After finishing this work, we received a preprint by Asakawa, Heinz and
M\"uller\cite{Heinz} which addresses similar issues.

\end{multicols}


\begin{thebibliography}{10}

\bibitem{Bonometto:1993pj}
For instance, see
S.~A.~Bonometto and O.~Pantano,
Phys.\ Rept.\  {\bf 228}, 175 (1993) and references therein.

\bibitem{Blaizot:1999bv}
For instance, see
J.~Blaizot,
``Signals of the quark-gluon plasma in nucleus nucleus collisions,''
for QM 99, hep-ph/9909434, and references therein.

\bibitem{Satz:1998bd}
For instance, see 
H.~Satz,
``A brief history of $J/\psi$ suppression,''
hep-ph/9806319,
and references therein.

\bibitem{Gale:1999ve}
C.~Gale, S.~Jeon and J.~Kapusta,
``Coherence time effects on $J/\psi$ production and suppression in
relativistic heavy ion collisions,''
hep-ph/9912213.

\bibitem{Jeon:1999gr}
S.~Jeon and V.~Koch,
Phys.\ Rev.\ Lett.\  {\bf 83} (1999) 5435.

\bibitem{Baym:1999up}
G.~Baym and H.~Heiselberg,
Phys.\ Lett.\  {\bf B469}, 7 (1999).

\bibitem{BJK} A manuscript detailing Ref.\protect\cite{Jeon:1999gr} and the
present one.

\bibitem{Bjorken:1983qr}
J.~D.~Bjorken,
Phys.\ Rev.\  {\bf D27}, 140 (1983).

\bibitem{Gottlieb:1997ae}
S.~Gottlieb {\it et al.},
Phys.\ Rev.\  {\bf D55}, 6852 (1997).

\bibitem{Whitmore:1976ip}
J.~Whitmore,
Phys.\ Rept.\  {\bf 27}, 187 (1976).

\bibitem{Gavin:1999bk}
S.~Gavin,
``Extraordinary baryon fluctuations and the QCD tricritical point,''
nucl-th/9908070.

\bibitem{Heinz}
M.Asakawa, U.Heinz and B.M\"uller, hep-ph/0003169.
\end{thebibliography}
\end{document}